\documentclass[11pt, a4paper]{article}
\pdfoutput=1 

\usepackage[height=8.85in,width=6.75in]{geometry}

\usepackage{graphicx,rotating}     
\usepackage[bookmarksopen,colorlinks=true,linkcolor=steelblue,
citecolor=orangered,urlcolor=darkred,linktoc=all]{hyperref}

\usepackage{amsmath,amssymb}
\usepackage{slashed}
\usepackage{bm}
\usepackage{bbm}
\usepackage{cite}
\usepackage{comment}
\usepackage[utf8]{inputenc}
\usepackage{accents}
\usepackage[footnotesize]{caption}
\usepackage{cancel}

\newcommand{\abs}[1]{\left\vert {#1} \right\vert}

\makeatletter
\@addtoreset{equation}{section}

\makeatother

\usepackage{xcolor}
\definecolor{darkred}{rgb}{0.7, 0., 0.}
\definecolor{orangered}{rgb}{1,0.27,0.}
\definecolor{steelblue}{rgb}{0.275,0.51, 0.706}
\definecolor{forestgreen}{rgb}{0.13,0.55,0.13}

\usepackage{tikz}
\usepackage{tikz-feynman}
\tikzfeynmanset{compat=1.0.0}
\pgfdeclarelayer{bg}    
\pgfsetlayers{bg,main}  

\begin{document}

\hypersetup{pageanchor=false}
\begin{titlepage}

\begin{center}

\hfill UMN-TH-4123/22 \\
\hfill FTPI-MINN-22-14 \\

\vskip 0.5in

{\Huge \bfseries Improved indirect limits on \vspace{5mm} \\ charm and bottom quark EDMs
} \\
\vskip .8in

{\Large Yohei Ema$^{a}$, Ting Gao$^{b}$, Maxim Pospelov$^{a,b}$}

\vskip .3in
\begin{tabular}{ll}
$^a$& \!\!\!\!\!\emph{William I. Fine Theoretical Physics Institute, School of Physics and Astronomy,}\\[-.3em]
& \!\!\!\!\!\emph{University of Minnesota, Minneapolis, M 55455, USA} \\
$^b$& \!\!\!\!\!\emph{School of Physics and Astronomy, University of Minnesota, Minneapolis, MN 55455, USA}
\end{tabular}

\end{center}
\vskip .6in

\begin{abstract}
\noindent
We derive indirect limits on the charm and bottom quark electric dipole moments (EDMs) from
paramagnetic AMO and neutron EDM experiments. 
The charm and bottom quark EDMs generate $CP$-odd photon-gluon operators and light quark EDMs
at the $c$- and $b$-quark mass thresholds. These $CP$-odd operators induce the $CP$-odd semi-leptonic operator $C_S$ 
and the neutron EDM below the QCD scale
that are probed by the paramagnetic and neutron EDM experiments, respectively. 
The bound from $C_S$ is $\vert d_c \vert < 1.3\times 10^{-20}\,e\,\mathrm{cm}$ 
for the charm quark and $\vert d_b \vert < 7.6\times 10^{-19}\,e\,\mathrm{cm}$ for the bottom quark,
with its uncertainty estimated as 10\,\%.
The neutron EDM provides a stronger bound, $\vert d_c \vert < 6\times 10^{-22}\,e\,\mathrm{cm}$
and $\vert d_b \vert < 2\times 10^{-20}\,e\,\mathrm{cm}$,
though with a larger hadronic uncertainty.
\end{abstract}

\end{titlepage}

\tableofcontents
\renewcommand{\thepage}{\arabic{page}}
\renewcommand{\thefootnote}{$\natural$\arabic{footnote}}
\setcounter{footnote}{0}
\hypersetup{pageanchor=true}

\section{Introduction}
\label{sec:intro}
Searches for electric dipole moments (EDMs) play a crucial role in probing physics beyond the Standard Model (SM).
Given the smallness of the SM contribution from the Cabbibo-Kobayashi-Maskawa phase,
EDMs of elementary particles are ideal observables to look for new physics that violates $CP$ symmetry
(see~\cite{Pospelov:2005pr} for a review). 

EDMs of elementary particles may be classified into three categories: 
lepton EDMs, light quark EDMs and heavy quark EDMs,
where ``light" and ``heavy" are in comparison with the QCD scale.
Among the lepton EDMs, the electron EDM is tested by paramagnetic EDM experiments
such as the ACME experiments~\cite{ACME:2018yjb}, while the muon and tau EDMs can be probed indirectly 
through paramagnetic and diamagnetic EDM experiments~\cite{Grozin:2008nw,Graner:2016ses,Ema:2021jds}.
Motivated by the muon $g-2$ anomaly~\cite{Aoyama:2020ynm,Muong-2:2021ojo}, 
the muon EDM has been and will be measured by the storage ring experiments 
as well~\cite{Muong-2:2008ebm,Semertzidis:1999kv,Iinuma:2016zfu,Abe:2019thb,Adelmann:2021udj}.
Hadronic EDMs can be induced by the EDMs of light quarks, 
as well as by other operators such as chromo-EDMs (CEDMs) and three-gluon $CP$-odd operators. 
The light quark EDMs, in particular the up and down quark EDMs, contribute to the neutron EDM
and are constrained by the neutron EDM experiments~\cite{Purcell:1950zz,Abel:2020pzs}.
Discussion of the strange quark contribution was typically focused on the color EDM contributions 
\cite{Khatsimovsky:1987bb,Fuyuto:2013gla}, while no reliable estimate of the neutron EDM induced by the strange quark exists at this point. 
Among the heavy quark EDMs, the top quark EDM is arguably most deeply investigated.
In particular, due to its large Yukawa coupling, the top quark EDM generates light fermion EDMs
through diagrams with an intermediate Higgs, which allows one 
to indirectly constrain the top quark EDM~\cite{Cirigliano:2016njn,Cirigliano:2016nyn}.
The light quark CEDMs directly contribute to the neutron EDM,
while the heavy quark CEDMs generate, \emph{e.g.}, 
the three-gluon Weinberg operator at the threshold, which induce the neutron EDM after the 
confinement~\cite{Weinberg:1989dx,Chang:1990jv,Boyd:1990bx,Dine:1990pf,Demir:2002gg,Sala:2013osa}.

Direct measurements of the heavy quark EDMs are difficult due to their short lifetimes.
The current strongest direct bound is from $e^+e^- \to q \bar{q}$ at LEP
and is only of the order of $10^{-17}\,e\,\mathrm{cm}$~\cite{Blinov:2008mu}.
Recently, an LHC based experiment is proposed that aims at directly measuring charm baryon dipole moments~\cite{Baryshevsky:2016cul,Botella:2016ksl,Fomin:2017ltw,Bagli:2017foe,Aiola:2020yam}, potentially improving the direct bounds on the heavy quark EDMs.
Given this situation, in this paper, we address indirect limits on the heavy quark EDMs.
The heavy quark EDMs induce several $CP$-odd operators below the heavy quark mass scale.
These operators generate observable signals in neutron/atomic/molecule EDM experiments,
which allows us to impose indirect limits on the heavy quark EDMs.
Thus, our goal of this paper is to understand the current status of indirect limits on the heavy quark EDMs.

Indirect limits on the charm and bottom quark EDMs were previously considered in~\cite{Gisbert:2019ftm} and the CEDM case was also analyzed in \cite{Haisch:2021hcg}.
There the authors derived limits based on the constraints on the heavy quark CEDMs.
Indeed, the charm and bottom quark EDMs well above the heavy quark mass scales, say 1\,TeV, 
generate the CEDM operators through the renormalization group (RG) running at a lower energy scale.
These CEDMs in turn source the three-gluon Weinberg operator
after integrating out the charm and bottom quark.
The Weinberg operator then generates the neutron EDM,
and this allows one to derive limits on the charm and bottom quark EDMs.
We may phrase it a re-interpretation of bounds on 
the heavy quark CEDMs at the heavy quark mass threshold,
with an assumption that there is no cancellation between the EDM
and CEDM contributions (see also discussion at the end of Sec.~\ref{sec:CS}).
In contrast, in this paper, we study 
$CP$-odd operators generated from the EDM operators at the heavy quark mass threshold.
Therefore our consideration directly applies to the heavy quark EDMs at the quark mass threshold and
is independent of~\cite{Gisbert:2019ftm}.
Even though the $CP$-odd operators induced by the heavy quark EDMs are formally higher dimensional,
suppressions from the charm and bottom quark masses are not severe as there is only a little hierarchy
between the quark masses and the QCD scale.

In order to make our limits robust, 
we derive indirect limits on charm and bottom quark EDMs 
based on multiple observables, paramagnetic and neutron EDMs.
Indirect limits always have a potential of having a cancellation among different operators.
For instance, paramagnetic EDM experiments
are sensitive to only a particular linear combination 
of the electron EDM $d_e$ and the $CP$-odd semi-leptonic operator $C_S$ (see Sec.~\ref{sec:CS}).
Therefore a new physics contribution to $d_e$ and $C_S$ can be such that the linear contribution
almost vanishes even though each term is sizable.
Deriving constraints from two completely different observables,
paramagnetic and neutron EDMs, makes the probability of having such a cancellation less likely.
We also minimize the QCD uncertainty as much as possible.
For this purpose the paramagnetic EDM plays an essential role
as it is sensitive to the heavy quark EDM through the semi-leptonic operator $C_S$.
The estimation of $C_S$ is not polluted by hadronic uncertainties, and
its uncertainty is theoretically well under control.

The structure of this paper is as follows. In Sec.~\ref{sec:aboveQCD},
we derive $CP$-odd operators generated by a heavy quark EDM after integrating out the heavy quark.
In particular, we see that the heavy quark EDM generates $CP$-odd photon-gluon operators and light quark EDMs.
We restrict ourselves above the QCD scale in this section, and thus the operators are written
in terms of light quarks and gluons.
In Sec.~\ref{sec:CS}, we calculate the semi-leptonic $CP$-odd operator $C_S$ induced 
from the $CP$-odd photon-gluon operator below the QCD scale.
This allows us to derive a clean constraint on the heavy quark EDMs from the paramagnetic EDM experiments.
Sec.~\ref{sec:neutronEDM} is devoted to an estimation of the neutron EDM induced by the $CP$-odd photon-gluon
operators and the light quark EDMs.
There we derive limits on the heavy quark EDMs from the neutron EDM experiments,
which are stronger than the paramagnetic EDM experiments but have a larger hadronic uncertainties.
The conventions of this paper is summarized in App.~\ref{app:convention},
while some technical details of our computation can be found in App.~\ref{app:details}.

\section{$CP$-odd operators from heavy quark EDMs}
\label{sec:aboveQCD}
In this section, we compute $CP$-odd operators induced by integrating out
charm and bottom quarks with EDMs.
The Lagrangian is given by
\begin{align}
	\mathcal{L} = \bar{Q} \left[i \slashed{D} - m_Q
	- \frac{id_Q}{2} \sigma_{\mu\nu} \gamma_5 F^{\mu\nu}\right] Q
	- \frac{1}{4}F^{\mu\nu}F_{\mu\nu}
	- \frac{1}{4}G^{a\mu\nu}G_{\mu\nu}^a,
\end{align}
where $Q$ is the heavy quark
with its mass $m_Q$ and its EDM $d_Q$.
The covariant derivative is defined by
\begin{align}
		iD_\mu &= i \partial_\mu + g_s G_\mu - e Q_Q A_\mu,
		\quad
		G_\mu = T^a G_\mu^a,
\end{align}
where $G_\mu^a$ is the gluon field with its field strength $G_{\mu\nu}^a$ and coupling $g_s$,
and $A_\mu$ is the photon field with its field strength $F_{\mu\nu}$ and coupling $e$.
The SU(3) generator is denoted by $T^a$ while the U(1) charge by $Q_Q$,
with $Q_c = Q_t = +2/3$ and $Q_b = -1/3$.
See App.~\ref{app:convention} for more details on our conventions.
Throughout this paper, ``heavy" means that the mass is larger than the QCD scale.
Although our discussion applies equally to the top quark, it has another contribution
that puts a stronger constraint on the EDM~\cite{Cirigliano:2016njn,Cirigliano:2016nyn}. 
Therefore our main focus is on the charm and bottom quarks, \emph{i.e.}, $Q = c, b$.

Once we integrate out the heavy quark, $d_Q$ generates several $CP$-odd operators.
For our purpose, there are two types of operators that are important: 
$CP$-odd photon-gluon operators and light quark EDMs.
Below the QCD scale, the former generates the neutron EDM and the semi-leptonic $CP$-odd interaction $C_S$ that induces paramagnetic EDMs, while the latter contributes to the neutron EDM.
In this section, we keep ourselves above the QCD scale.
The operators are then written in terms of the quarks and gluons. 
We will deal with nonperturbative effects at the QCD scale in the subsequent sections.

\subsection{$CP$-odd photon-gluon operators}
\label{subsec:photon-gluon}

\begin{figure}[t]
	\centering
 	\includegraphics[width=0.55\linewidth]{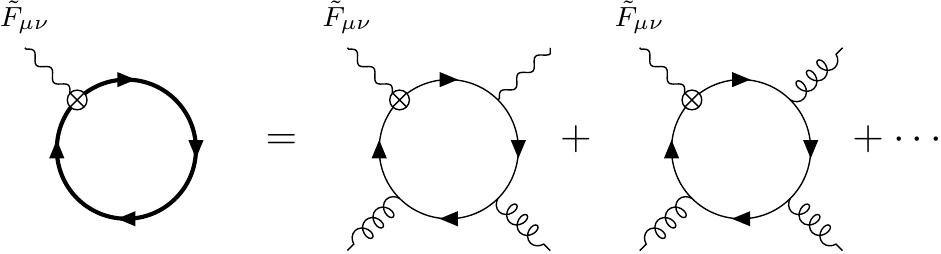}
	\caption{\small The effective action to linear order in $d_Q$ after integrating out the heavy quark.
	The thick line on the left hand side indicates
	 the full heavy quark propagator with the covariant derivative $D_\mu$,
	and the cross dot indicates the EDM operator insertion.
	The full propagator is expanded with respect to the field strengths,
	which results in the $CP$-odd photon-gluon operators
	as shown on the right hand side.}
	\label{fig:photon_gluon}
\end{figure}

The heavy quark EDM induces $CP$-odd photon-gluon operators 
through the diagrams in Fig.~\ref{fig:photon_gluon}.
The effective action after integrating out the heavy quark,
to linear order in $d_Q$, is given by
\begin{align}
	S_\mathrm{eff} = - \frac{i d_Q}{2}\mathrm{Tr}
	\left[\frac{1}{i\slashed{D} - m_Q}\sigma_{\mu\nu} \tilde{F}^{\mu\nu}\right],
	\label{eq:effective_action_dQ}
\end{align}
where the trace is taken over the spinor, the color and the spacetime.
We further expand this expression with respect to the field strength $H_{\mu\nu}$, given by
\begin{align}
	H_{\mu\nu} = g_s G_{\mu\nu} -eQ_Q F_{\mu\nu},
\end{align}
which satisfies $[i D_\mu, i D_\nu] = iH_{\mu\nu}$.
To the lowest order, we obtain 
(see App.~\ref{app:photon-gluon} for derivation)
\begin{align}
	\mathcal{L}_\mathrm{eff} = \frac{d_Q}{48\pi^2 m_Q^3}
	\mathrm{tr}_c \left[- H_{\mu\nu}H^{\mu\nu} H_{\rho\sigma}\tilde{F}^{\rho\sigma}
	+ 2 {H^\mu}_\nu {H^\nu}_\rho {H^\rho}_\sigma {\tilde{F}^\sigma}_\mu
	\right],
	\label{eq:Leff_photon-gluon}
\end{align}
where $\mathrm{tr}_c$ is the trace only over the color.
This contains $CP$-odd photon-gluon operators as well as a $CP$-odd pure photon operator.
The operator quadratic in the photon field is given by
\begin{align}
\label{eq:2gl-2ga}
	\mathcal{L}_{G^2F\tilde{F}} &= \frac{eQ_Q g_s^2 d_Q}{24\pi^2 m_Q^3}
	\mathrm{tr}_c \left[
	F_{\mu\nu}G^{\mu\nu} \tilde{F}_{\rho\sigma}G^{\rho\sigma}
	-{\tilde{F}^\mu}_\nu {G^\nu}_\rho {F^\rho}_\sigma {G^\sigma}_\mu
	\right],
\end{align}
while the one linear in the photon field is given by
\begin{align}
\label{eq:3gl-1ga}
	\mathcal{L}_{G^3\tilde{F}} &= \frac{g_s^3 d_Q}{48\pi^2 m_Q^3}
	\mathrm{tr}_c \left[- G_{\mu\nu}G^{\mu\nu} G_{\rho\sigma}\tilde{F}^{\rho\sigma}
	+ 2 {G^\mu}_\nu {G^\nu}_\rho {G^\rho}_\sigma {\tilde{F}^\sigma}_\mu
	\right].
\end{align}
Carrying out color traces explicitly, one finds that (\ref{eq:2gl-2ga}) contains $\delta^{ab}$, while  (\ref{eq:3gl-1ga}) has $d^{abc}$ structure. In that sense, (\ref{eq:2gl-2ga}) would exist for any choice of the gauge group for $G_{\mu\nu}$ including a $\mathrm{U}(1)$, while (\ref{eq:3gl-1ga}) requires $N\geq 3$ for $ \text{SU}(N) $, which includes of course the color group of the Standard Model. 

As we will see, below the QCD scale,
operator (\ref{eq:2gl-2ga}) contributes to paramagnetic EDMs through the semi-leptonic $CP$-odd operator $C_S$
while (\ref{eq:3gl-1ga}) contributes to the neutron EDM.
The implication of the $CP$-odd pure photon operator is studied in detail in~\cite{Ema:2021jds} 
in the context of the muon EDM.
This operator puts only a subdominant constraint on the heavy quark EDM,
and thus we do not discuss it any further in this paper.

\subsection{Light quark EDMs}
\label{subsec:udEDMs}

\begin{figure}[t]
	\centering
 	\includegraphics[width=0.3\linewidth]{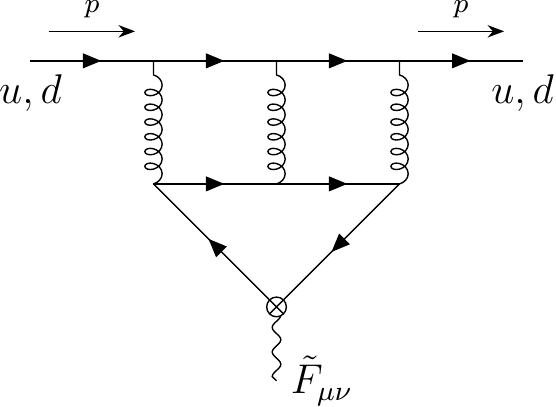}
	\caption{\small An example of the three-loop diagrams that generate the light quark EDMs.
	The cross dot indicates the heavy quark EDM $d_Q$ insertion, the wavy line is the external photon, 
	the closed solid line is the heavy quark and the upper solid line is the light quark, respectively.
	There are five additional diagrams that are permutations of the gluon lines attached
	to the light quark line. }
	\label{fig:three_loop}
\end{figure}

The heavy quark EDM induces the up and down quark EDMs at 
the three-loop level through the diagram
in Fig.~\ref{fig:three_loop} (and its permutations),
where the closed loop is the heavy quark and the cross dot indicates the heavy quark EDM insertion. 
In our case the heavy quark EDM operator already contains a derivative acting on the external photon,
and thus we can ignore the momentum flow of the photon.
The mass and the external momentum of the light quark are small compared to the heavy quark mass scale,
and this allows us to expand the diagram with respect to the light quark mass and momentum.
To linear order, the amplitude is schematically given by
\begin{align}
	i\mathcal{M}_{3\mathchar`-\mathrm{loop}} = i \tilde{F}^{\mu\nu}p^\rho \bar{q} S_{\mu\nu\rho} q,
\end{align}
where $q = u, d$ is the light quark field and $p$ is the light quark momentum.
Here we used the equation of motion to express quantities that explicitly depend on the light quark mass $m_q$ in terms of $p$.
After projecting $S_{\mu\nu\rho}$ onto the EDM part, we obtain the light quark EDM $d_q$ as
\begin{align}
	d_q = \frac{m_q}{96}\mathrm{tr}\left[S_{\mu\nu\rho}\left\{\sigma^{\mu\nu},\gamma^\rho\right\}\right],
\end{align}
where the trace is taken only over the spinor index, and $\{..,..\}$ represents anticommutator.
We evaluate this expression with the dimensional regularization.
Since we expanded the amplitude with respect to the external momentum and the light quark mass,
$S_{\mu\nu\rho}$ depends only on the heavy quark mass and hence is given by the vacuum integrals.
In our case, among six propagators, three are massive and three are massless. 
In this case, we can reduce the vacuum integrals to two irreducible master integrals~\cite{Broadhurst:1991fi}.
We use \texttt{FIRE6}~\cite{Smirnov:2019qkx} to reduce the vacuum integrals by integration by parts,
and use analytical expressions of the master integrals in~\cite{Martin:2016bgz}.
We then obtain
\begin{align}
	d_q &= \frac{5\left(8\zeta(3)-7\right)}{72} \times \left(\frac{\alpha_s}{\pi}\right)^3\frac{m_q}{m_Q}d_Q,
\end{align}
where $\alpha_s = g_s^2/4\pi$ and $\zeta(3) \simeq 1.202$ with $\zeta(z)$ the Riemann zeta function.
We have checked that all the gauge-dependent part of the gluon propagator cancels in our computation.
Moreover, although each master integral is divergent, all the divergences cancel and the final result is finite.
If we drop the color factors, this is equivalent to one contribution to the electron EDM from
the muon/tau EDM, the diagram~(a) in~\cite{Grozin:2008nw}.
Our result differs from~\cite{Grozin:2008nw} and 
we will come back to the origin of this difference in a separate publication~\cite{Ema:2022}.
Notice that $d_q$ is suppressed only by $1/m_Q$ since the loop integral is 
dominated by the loop momentum of order $m_Q$, while the $CP$-odd photon-gluon operators
that we computed in Sec.~\ref{subsec:photon-gluon} are suppressed by $1/m_Q^3$.
Therefore the light quark EDM is expected to be more important for heavier quarks,
although there is another (and a dominant) contribution in the case of the top quark as
we mentioned at the beginning of this section.

\section{Paramagnetic EDM}
\label{sec:CS}
In this section we study a constraint on the heavy quark EDM from 
paramagnetic atomic/molecule EDM experiments, in particular the ACME experiment~\cite{ACME:2018yjb}.
The paramagnetic EDM experiments are sensitive only to a linear combination of two operators,
the electron EDM $d_e$ and the semi-leptonic $CP$-odd operator $C_S$ given by
\begin{align}
	\mathcal{L} = C_S \frac{G_F}{\sqrt{2}}\bar{e}i\gamma_5 e \left(\bar{p}p + \bar{n}n\right),
\end{align}
where $G_F$ is the Fermi constant, $e$ is the electron (one should not confuse it with the electromagnetic coupling), 
$p$ is the proton and $n$ is the neutron, respectively.
It is convenient to define an \emph{equivalent} electron EDM as~\cite{Pospelov:2013sca}
\begin{align}
	d_e^{(\mathrm{equiv})} = d_e + C_S \times 1.5 \times 10^{-20}\,e\,\mathrm{cm},
\end{align}
where we focus on ThO molecule.
The current bound on the equivalent electron EDM is given by
$\vert d_e^{(\mathrm{equiv})}\vert < 1.1\times 10^{-29}\,e\,\mathrm{cm}$~\cite{ACME:2018yjb}.

\begin{figure}[t]
	\centering
 	\includegraphics[width=0.25\linewidth]{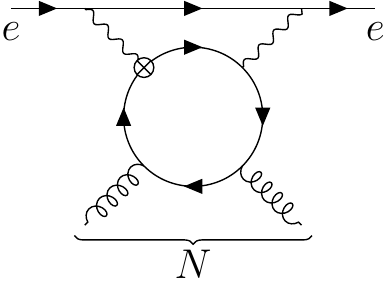}
	\caption{\small The diagram that generates the $CP$-odd semi-leptonic operator $C_S$.
	The photons are attached to the electron line and generate the structure $\bar{e}i\gamma_5 e$, 
	while the gluons feed into the nucleon $N$.}
	\label{fig:CS}
\end{figure}

We now compute $C_S$ induced by the $CP$-odd photon-gluon operator,
in particular the two-photon two-gluon operator, below the QCD scale.
The effective Lagrangian is given by
\begin{align}
	\mathcal{L}_{G^2 F\tilde{F}}= \frac{e Q_Q g_s^2 d_Q}{48\pi^2 m_Q^3} G_{\mu\nu}^a G^{a}_{\alpha\beta}
	\left(F^{\mu\nu}\tilde{F}^{\alpha\beta} - F^{\nu\alpha} \tilde{F}^{\beta\mu}\right).
\end{align}
Because our starting point here is explicitly isospin symmetric, 
it will result in the same $C_S$ coupling for neutrons and protons. 
The nucleon matrix element of the gluon part is given by
\begin{align}
	\langle N \rvert \frac{g_s^2}{32\pi^2}G^{a}_{\mu\nu}G^{a}_{\alpha\beta}\lvert N\rangle
	= -\frac{m_N}{108} \left(\eta_{\alpha\mu}\eta_{\beta \nu} - \eta_{\alpha\nu} \eta_{\beta\mu}\right)
	\bar{N}N + \cdots,
\end{align}
where $N$ is the nucleon field (either $p$ or $n$) with $m_N$ its mass,
and $\cdots$ denotes the traceless tensor part that is irrelevant for our purpose.
Here we used that, in the chiral limit, the one-loop trace anomaly dominantly contributes to the nucleon mass,
\begin{align}
	\langle N \rvert \frac{g_s^2}{32\pi^2} G_{\mu\nu}^a G^{a\mu\nu}\lvert N \rangle
	= - \frac{m_N}{9} \bar{N}N,
	\label{eq:GG_trace_anomaly}
\end{align}
where the coefficient in the right-hand-side is related to the beta function~\cite{Shifman:1978zn}.
The photons are attached to the electron and induce the operator $\bar{e}i\gamma_5 e$ 
at one-loop~\cite{Flambaum:2019ejc} 
as shown in Fig.~\ref{fig:CS}.
This integral is logarithmically divergent, which is regulated by the heavy quark mass scale.
Therefore, to the leading logarithmic accuracy, we obtain

\begin{align}
	C_S\frac{G_F}{\sqrt{2}} = - \frac{4 Q_Q \alpha^2}{27} \frac{m_N m_e}{m_Q^3} 
	\log \left(\frac{m_Q}{m_e}\right) \frac{d_Q}{e},
\end{align}
where $\alpha = e^2/4\pi$ and $m_e$ is the electron mass.
By requiring $\vert{d_e^{(\mathrm{equiv})}}\vert < 1.1 \times 10^{-29}\,e\,\mathrm{cm}$, we obtain
\begin{align}
	\abs{d_c} < 1.3\times 10^{-20}\,e\,\mathrm{cm},
\end{align}
for the charm quark and
\begin{align}
	\abs{d_b} < 7.6\times 10^{-19}\,e\,\mathrm{cm},
\end{align}
for the bottom quark, respectively.
The heavy quark EDM also induces the electron EDM at three-loop~\cite{Grozin:2008nw,Grozin:2009jq},
but its contribution to $d_e^{(\mathrm{equiv})}$ is negligible compared to $C_S$.

We now estimate the precision of our calculation.
There is an uncertainty associated with Eq.~\eqref{eq:GG_trace_anomaly},
which is valid only in the chiral limit.
However, the quark contribution to the nucleon mass is less than 10\,\%~\cite{Ellis:2018dmb},
and hence the uncertainty of Eq.~\eqref{eq:GG_trace_anomaly} is also less than 10\,\%.
Thanks to the lattice computations, these contributions are rather precisely known
and the uncertainty in the non-perturbative matrix element can be further reduced
by including the light quark contributions into our computation.
Another uncertainty originates from our photon-loop computation, where we include
only the leading logarithmic term.
The uncertainty associated with this treatment 
may be estimated by changing the cut-off scale from $m_Q$ to $2m_Q$, 
which results in $\lesssim 10\,\%$ change of the result for both the charm and bottom quarks.
This can be further improved by performing the full two-loop computation, 
with the heavy quark and photon loops at the same time.
Therefore,  the precision of our $C_S$ calculation is estimated to be 10\,\%
and this can be further improved by properly including the quark contribution to the nucleon mass
and performing the full two-loop computation of the photon-loop.

Our constraint is stronger than the one derived from $e^+ e^- \to q\bar{q}$ at LEP 
by several orders of magnitude~\cite{Blinov:2008mu}.
Although our constraint is weaker than~\cite{Gisbert:2019ftm} at its face value, there are two caveats.
First, as we mentioned in the introduciton, the constraint in~\cite{Gisbert:2019ftm} is a re-interpretation
of the bounds on the CEDMs $\tilde{d}_Q$ at the heavy quark mass scale, \emph{i.e.}, $\tilde{d}_{Q}(m_Q)$.
Both the quark EDM and CEDM at high energy scale $\Lambda_\mathrm{NP}$ contribute to the CEDM
at lower energy due to the RG running,
and hence $\tilde{d}_Q(m_Q)$ is given as a linear combination of $d_Q(\Lambda_\mathrm{NP})$
and $\tilde{d}_Q(\Lambda_\mathrm{NP})$.
With only this information, one can put a constraint on only this particular linear combination of
$d_Q(\Lambda_\mathrm{NP})$ and $\tilde{d}_Q(\Lambda_\mathrm{NP})$.
Therefore the authors assume that there is no huge cancellation between $d_Q(\Lambda_\mathrm{NP})$ and $\tilde{d}_Q(\Lambda_\mathrm{NP})$, which allow them to derive a constraint solely on $d_Q(\Lambda_\mathrm{NP})$
and thus on $d_Q(m_Q)$.
Since our constraint directly applies to $d_Q(m_Q)$, it is an independent constraint,
without any assumptions on the relative size of 
$d_Q(\Lambda_\mathrm{NP})$ and $\tilde{d}_Q(\Lambda_\mathrm{NP})$.
Second, it is a complicated task to estimate the size of the neutron EDM induced by the Weinberg operator.
Since~\cite{Gisbert:2019ftm} relies on the constraints on $\tilde{d}_Q(m_Q)$ from the Weinberg operator,
it has a large hadronic uncertainty.
In contrast, as we have just emphasized, 
our computation of $C_S$ is precise and its uncertainty is well under control.
Given the rapid progress of the paramagnetic EDM experiments,
$C_S$ will continue providing important and clean constraints on the heavy quark EDMs.
In the next section, we will see that the neutron EDM provides a stronger constraint than $C_S$,
comparable to~\cite{Gisbert:2019ftm}, though with a larger hadronic uncertainty.

\section{Neutron EDM}
\label{sec:neutronEDM}
In Sec.~\ref{sec:aboveQCD} we have seen that the heavy quark EDM
generates the $CP$-odd photon-gluon operator and the light quark EDM.
Below the QCD scale, these $CP$-odd operators in turn generate the neutron EDM $d_n$,
whose experimental upper bound is~\cite{Abel:2020pzs}\footnote{
	The $^{199}$Hg EDM experiment puts a similar constraint on the neutron EDM~\cite{Graner:2016ses}.
}
\begin{align}
	\abs{d_n} < 1.8\times 10^{-26}\,e\,\mathrm{cm}.
\end{align}
In this section we translate this upper bound to the bound on the heavy quark EDM
by estimating the size of the induced neutron EDM, with effects of the nonperturbative QCD taken into account (to the extent it is possible).

\subsection{Light quark EDM contribution}
\label{subsec:udEDM2nEDM}
We first study the neutron EDM induced by the light quark EDM $d_n^{(d_q)}$.
Both the QCD sum rule and the quark model suggest that~\cite{Pospelov:2000bw,Pospelov:2005pr,Hisano:2012sc}
\begin{align}
	d_n^{(d_q)} &= \frac{1}{3}\left(4d_d - d_u\right).
\end{align}
QCD sum rule calculations \cite{Pospelov:2000bw,Hisano:2012sc} and more recently lattice calculations \cite{Bhattacharya:2016zcn} give support to this simple formula within $\sim 30\%$ accuracy. 
We thus obtain
\begin{align}
	d_n^{(d_q)} &= \frac{5(8\zeta(3)-7)}{72}\left(\frac{\alpha_s}{\pi}\right)^3 
	\frac{4m_d - m_u}{3m_Q} d_Q.
\end{align}
As we will see, there is another contribution to the neutron EDM from the $CP$-odd photon-gluon operator
that leads to the  QCD condensate power corrections in the light quark propagator.
The rest of this section is devoted to estimate the size of this contribution.

\subsection{$CP$-odd photon-gluon operator contribution}
\label{subsec:QCDcondensate}
In Sec.~\ref{subsec:photon-gluon} we saw that the heavy quark EDM generates the $CP$-odd photon-gluon operator
\begin{align}
	\mathcal{L}_{G^3\tilde{F}} &= \frac{g_s^3 d_Q}{48\pi^2 m_Q^3}
	\mathrm{tr}_c \left[- G_{\mu\nu}G^{\mu\nu} G_{\rho\sigma}\tilde{F}^{\rho\sigma}
	+ 2 {G^\mu}_\nu {G^\nu}_\rho {G^\rho}_\sigma {\tilde{F}^\sigma}_\mu
	\right].
	\label{eq:photon-gluon-QCDsum}
\end{align}
Below the QCD scale the gluons confine and condense, and this operator feeds into the neutron EDM.
We use the QCD sum rule technique~\cite{Shifman:1978bx} 
to estimate the size of the neutron EDM induced by this operator.

The starting point of the QCD sum rule is to define the two-point correlator
\begin{align}
	\Pi(p) &\equiv i \int d^4x e^{ip\cdot x}\langle 0 \vert \mathcal{T}\left\{ \eta(x), \bar{\eta}(0)\right\} \vert 0\rangle.
\end{align}
The interpolating function $\eta$ is typically chosen as
\begin{align}
	\eta(x) = j_1(x) + \beta j_2(x),
\end{align}
and has an overlap with the neutron one-particle state.
The functions $j_1$ and $j_2$ have the same quantum numbers as the neutron and are given by
\begin{align}
	j_1 &= 2\epsilon_{ijk}\left(d_i^{T} \mathcal{C} \gamma_5 u_j\right) d_k,
	\quad
	j_2 = 2\epsilon_{ijk}\left(d_i^{T} \mathcal{C} u_j\right) \gamma_5d_k,
\end{align}
where $i, j, k$ are the color indices with $\epsilon_{ijk}$ the totally anti-symmetric tensor,
and $\mathcal{C}$ is the charge conjugate matrix 
(see App.~\ref{app:convention} for its definition).
Finally $\beta$ is a redundant parameter that should disappear in an exact computation. 
In practice, one can choose $\beta$, \emph{e.g.}, to
optimize the convergence of the operator production expansion (OPE).
The QCD sum rule relies on the quark-hadron duality and evaluate this two point correlator $\Pi(p)$
in the hadronic (phenomenological) side and the quark (OPE) side.
On the phenomenological side, $\Pi(p)$ is expressed
in terms of hadronic quantities such as the neutron mass and dipole moments.
On the OPE side, $\Pi(p)$ is evaluated in terms of perturbative quarks and gluons with non-perturbative effects included 
in the form of the QCD condensates.
One then obtains an estimation of the hadronic quantities by equating these two expressions,
after the Borel transformation to reduce effects of excited states.
In our case, we have external electromagnetic field $F_{\mu\nu}$ multiplying three Lorentz structures that in principle we can use to derive the sum rules:
$\slashed{p}\sigma_{\mu\nu} \slashed{p}$, $\{\slashed{p}, \sigma_{\mu\nu}\}$ and $\sigma_{\mu\nu}$.
In this paper, we focus on the sum rule that follows from $\slashed{p}\sigma_{\mu\nu} \slashed{p}$
since it depends most strongly on the momentum and the vacuum susceptibilities are less relevant~\cite{He:1996wy}.
Moreover, it is computationally simple to derive the QCD sum rule based on this structure.
In the following we evaluate $\Pi(p)$ in the OPE and phenomenological sides, respectively.

\paragraph{OPE side.}
We denote the light quark propagator as
\begin{align}
	\left\langle 0 \vert \mathrm{T}\left\{q_{i}(x), \bar{q}_{j}(y)\right\} \vert 0 \right\rangle 
	\equiv
	\delta_{ij} S(x, y),
\end{align}
where only the color-diagonal contribution is relevant for the QCD sum rule
based on $\slashed{p}\sigma_{\mu\nu} \slashed{p}$ in our case.
The $CP$-odd photon-gluon operator does not distinguish the up and down quarks,
and thus we collectively denote $u$ and $d$ as $q$.
With this expression, we obtain
\begin{align}
	\Pi_\mathrm{OPE}(p) = -24 i\int d^4x e^{ip\cdot x}&\left\{
	\mathrm{tr}\left[\gamma_5 S^c \gamma_5 S\right] S + S \gamma_5 S^c \gamma_5 S
	+ \beta^2\left(\gamma_5 S S^c S \gamma_5 + \mathrm{tr}\left[S^c S\right] \gamma_5 S \gamma_5 \right)
	\right. \nonumber \\ &\left.
	+\beta\left(S S^c \gamma_5 S \gamma_5 + \gamma_5 S \gamma_5 S^c S
	+ \mathrm{tr}\left[S S^c \gamma_5\right] S \gamma_5
	+ \mathrm{tr}\left[S^c S \gamma_5\right] \gamma_5 S
	\right)
	\right\},
	\label{eq:correlator_expand}
\end{align}
where 
$S^{c}_{\alpha\beta} \equiv \left(\mathcal{C} S^T \mathcal{C}\right)_{\alpha\beta}$
and  $S^{T}_{\alpha\beta} \equiv S_{\beta\alpha}$
with $\alpha$ and $\beta$ being the spinor indices,
and the trace is taken only over the spinor index.
In our case the light quark propagator has two contributions
\begin{align}
	S &= S^{(0)} + S^{(\cancel{{CP}})},
\end{align}
where $S^{(0)}$ corresponds to the free quark propagator and is given by
\begin{align}
	S^{(0)}(x) &= \frac{i \slashed{x}}{2\pi^2 x^4},
\end{align}
while $S^{(\cancel{{CP}})}$ is the $CP$-odd part which we compute from now.

The source inserted into the correlator has three gluon fields, and we follow the general idea of \cite{Demir:2002gg} where the Weinberg operator contribution to $d_n$ was first evaluated, and where two gluons are treated perturbatively, while the third gluon field contributes to the quark-gluon condensate.
With the QCD condensate background, we can write down the diagram
\begin{align}
	S_{ij}^{(\cancel{{CP}})}(p) = 
	\begin{tikzpicture}[baseline=(a)]
	\begin{feynman}[inline = (base.a), horizontal=a to c]
		\vertex [label=270:\({\scriptstyle j}\)] (a);
		\vertex [right = of a] (b);
		\vertex [right = of b] (c);
		\vertex [below = of b] (d);	
		\vertex [below =of c] (e);
		\vertex [above =of c] (f);
		\node [left = -0.1cm of c, crossed dot, fill=white] (g);
		\vertex [right =of c] (h);
		\vertex [below =of h] (i);
		\vertex [label=270:\({\scriptstyle i}\), right =of h] (j);
		\begin{pgfonlayer}{bg}
		\draw[draw=gray]  (e) ellipse (1.3 and 0.35);
		\diagram*{
		(a) -- [fermion, momentum=\(\scriptstyle p\)] (b),
		(b) -- [fermion] (d),
		(b) -- [insertion={1.0}] (d),
		(b) -- [gluon, momentum=\(\scriptstyle p\)] (c),
		(c) -- [gluon, insertion={1.0}] (e),
		(c) -- [photon, insertion={1.0}] (f),
		(c) -- [gluon, momentum=\(\scriptstyle p\)] (h),
		(i) -- [fermion] (h) -- [fermion, momentum=\(\scriptstyle p\)] (j),
		(i) -- [fermion, insertion={0.0}] (h),
		};
		\end{pgfonlayer}
	\end{feynman}
	\end{tikzpicture},
\end{align}
where the crosses indicate the background fields, either the external photon or the QCD condensation of
the quarks and gluons, and the cross dot is 
the insertion of the $CP$-odd photon-gluon operator~\eqref{eq:photon-gluon-QCDsum}.
Since the QCD vacuum does not violate the Lorentz and color symmetries,
we have
\begin{align}
	\left\langle q_{i} \bar{q}_{j} G^a_{\mu\nu}\right\rangle &= 
	-\frac{1}{192}T^{a}_{ij} \sigma_{\mu\nu} \left\langle \bar{q} \sigma \cdot G q\right\rangle,
	\label{eq:QCDcondensate}
\end{align}
where $\sigma \cdot G = \sigma_{\mu\nu}G^{\mu\nu}$,
and $\langle \cdots \rangle$ corresponds to the vacuum expectation value.
We thus obtain
\begin{align}
	S^{(\cancel{{CP}})}_{ij}(p) &= 
	\delta_{ij}S^{(\cancel{{CP}})}(p) 
	= i\delta_{ij} 
	\frac{5\alpha_s^2 d_Q}{864 m_Q^3}\frac{g_s \langle \bar{q} \sigma\cdot Gq\rangle}{p^2}
	\frac{i}{\slashed{p}}\left[\sigma\cdot \tilde{F} - \frac{2p^\mu p^\nu}{p^2}\tilde{F}_{\mu\alpha}{\sigma_\nu}^\alpha\right]
	\frac{i}{\slashed{p}}.
	\label{eq:light_quark_EDM_mom}
\end{align}
Its Fourier transformation has an IR divergence,
which in the dimensional regularization leads
\begin{align}
	S^{(\cancel{{CP}})}(x) 
	&= \frac{5 \alpha_s^2 d_Q}{27648\pi^2 m_Q^3} g_s \langle \bar{q}\sigma \cdot G q\rangle
	\Gamma(\epsilon_\mathrm{IR}) \left(-\Lambda^2_\mathrm{IR} x^2\right)^{-\epsilon_\mathrm{IR}}
	\tilde{F}\cdot \sigma,
\end{align}
where we take $d_\mathrm{IR} = 4+2\epsilon_\mathrm{IR}$ and keep only the logarithmic terms,
with $\Lambda_\mathrm{IR}$ the IR cut-off scale.
Due to the sensitivity to IR scale, calculations of further terms in the OPE are not possible. As a consequence, this evaluation should be viewed as an estimate, which cannot be systematically improved within QCD sum rule method. 
With the explicit forms of $S^{(0)}$ and $S^{(\cancel{{CP}})}$, we obtain
\begin{align}
	\Pi_\mathrm{OPE}(p) 
	= -\frac{5 \alpha_s^2 d_Q}{2^{12}3^3 \pi^4 m_Q^3}\left\langle \bar{q} \sigma \cdot g_s G q\right\rangle
	\left[\left(1-\beta\right)^2 \slashed{p} \tilde{F} \cdot \sigma \slashed{p} 
	- 18\left(1-\beta^2\right) p^2 \tilde{F}\cdot \sigma\right]
	I(p^2; \epsilon_\mathrm{IR}, \epsilon_{\mathrm{UV}}),
	\label{eq:Pi_OPE_bfrBorel}
\end{align}
where we perform the dimensional regularization to tame the UV divergence
with $d_\mathrm{UV} = 4-2\epsilon_{\mathrm{UV}}$, and
\begin{align}
	I(p^2; \epsilon_\mathrm{IR}, \epsilon_\mathrm{UV})
	&\equiv \Gamma(\epsilon_\mathrm{IR})
	\left(-\frac{\Lambda_\mathrm{IR}^2}{p^2}\right)^{-\epsilon_\mathrm{IR}}
	\Gamma(-\epsilon_\mathrm{UV}-\epsilon_{\mathrm{IR}}) 
	\left(-\frac{\mu^2}{p^2}\right)^{-\epsilon_\mathrm{UV}},
\end{align}
with $\mu$ the renormalization scale.
We then perform the Borel transformation, defined as~\cite{Ioffe:1983ju}
\begin{align}
	\mathcal{B}\left[\Pi(p^2 = -P^2)\right]
	&= \frac{1}{\pi}\int_0^\infty \frac{dP^2}{M^2}e^{-P^2/M^2}
	\mathrm{Im}\left[\Pi(p)\right]_{p^2=-P^2}.
\end{align}
We thus obtain the $\slashed{p}\sigma_{\mu\nu}\slashed{p}$ part as
\begin{align}
	\mathcal{B}\left[\Pi_\mathrm{OPE}(p^2 = -P^2)\right]_{\slashed{p} \tilde{F}\cdot \sigma \slashed{p}}
	&= -\frac{5\alpha_s^2 d_Q}{2^{12} 3^3 \pi^4 m_Q^3}\left(1-\beta\right)^2 
	\left\langle \bar{q} \sigma \cdot g_s G q\right\rangle \log \left(\frac{M^2}{\Lambda_\mathrm{IR}^2}\right).
	\label{eq:dn_Borel_OPE}
\end{align}
Taking the imaginary part of $I(p^2; \epsilon_\mathrm{IR}, \epsilon_\mathrm{UV})$ 
is somewhat subtle, and we provide details in App.~\ref{app:Borel_IRdiv}.
There we also clarify the physical meaning of the scale $\Lambda_\mathrm{IR}$;
it should be identified with the mass of the constituent quarks.
This expression is to be compared with the phenomenological expression.

\paragraph{Phenomenological side.}
On the phenomenological side, the two-point correlator with the neutron EDM insertion is expressed as
\begin{align}
	\Pi_\mathrm{pheno}(p) = -\lambda_n^2\left[\frac{\slashed{p}+m_n}{p^2 - m_n^2} 
	- \frac{d_n}{2(p^2-m_n^2)^2}\slashed{p}\tilde{F}\cdot \sigma \slashed{p} + \cdots \right],
\end{align}
where $m_n$ is the neutron mass and $\lambda_n$ parametrizes the overlap between the interpolating function
$\eta$ and the neutron one-particle state.
After the Borel transformation, we obtain
\begin{align}
	\mathcal{B}\left[\Pi_\mathrm{pheno}(p^2 = -P^2)\right]_{\slashed{p} \tilde{F}\cdot \sigma \slashed{p}}
	&= \frac{\lambda_n^2 d_n}{2M^4} e^{-m_n^2/M^2}
	+ \cdots,
	\label{eq:dn_Borel_pheno}
\end{align}
where $\cdots$ expresses contributions from excited states which we ignore in the following.

\paragraph{Sum rule.}
The QCD sum rule of the neutron EDM is obtained 
by equating Eqs.~\eqref{eq:dn_Borel_OPE} and~\eqref{eq:dn_Borel_pheno}.
We thus obtain
\begin{align}
	\frac{\lambda_n^2 d_n^{(G^3\tilde{F})}}{2M^4} e^{-m_n^2/M^2} &= 
	-\frac{5\alpha_s^2 d_Q}{2^{12} 3^3 \pi^4 m_Q^3}\left(1-\beta\right)^2 
	\left\langle \bar{\psi} \sigma \cdot g_s G \psi\right\rangle \log \left(\frac{M^2}{\Lambda_\mathrm{IR}^2}\right),
\end{align}
where $d_n^{(\tilde{F}G^3)}$ is the neutron EDM induced by the $CP$-odd photon-gluon operator (to distinguish it
from the one induced by the light quark EDM).
We may use the Ioffe formula for the nucleon mass~\cite{Ioffe:1981kw,Leinweber:1995fn}\footnote{
	The Ioffe formula in~\cite{Leinweber:1995fn}
	is derived based on $\chi_\mathrm{SR}/2$ which is equivalent to our $-\eta/2$ with $\beta=-1$.
	Therefore the normalization of $\lambda_n$ differs by a factor two.
}
\begin{align}
	\frac{\lambda_n^2 m_n}{M^4}e^{-m_n^2/M^2} = 
	-\frac{7-2\beta-5\beta^2}{16\pi^2}\langle \bar{q}q\rangle,
\end{align}
to eliminate $\lambda_n$.
Then we obtain the QCD sum rule estimation of the neutron EDM
\begin{align}
	d_n^{(G^3\tilde{F})} &= d_Q \times \frac{5\alpha_s^2}{2^7 3^3 \pi^2}
	\frac{m_n m_0^2}{m_Q^3}
	\frac{1-\beta}{7 + 5\beta}
	\log\left(\frac{M^2}{\Lambda_\mathrm{IR}^2}\right),
\end{align}
where we used $\langle q \sigma \cdot g_sGq\rangle = m_0^2 \langle \bar{q}q\rangle$
with $m_0^2 = 0.8\,\mathrm{GeV}^2$~\cite{Belyaev:1982sa}.

\subsection{Constraint on heavy quark EDM}
\label{subsec:constraint_dn}

The neutron EDM has two contributions, induced by the light quark EDM and the $CP$-odd photon-gluon operators,
and is given by
\begin{align}
	d_n = d_n^{(d_q)} + d_n^{(G^3\tilde{F})},
\end{align}
where
\begin{align}
	d_n^{(d_q)} &= d_Q \times \frac{5(8\zeta(3)-7)}{72}\left(\frac{\alpha_s}{\pi}\right)^3 
	\frac{4m_d - m_u}{3m_Q},
	\quad
	d_n^{(G^3\tilde{F})} = d_Q \times \frac{5\alpha_s^2}{2^7 3^3 \pi^2}
	\frac{m_n m_0^2}{m_Q^3}
	\log\left(\frac{M^2}{\Lambda_\mathrm{IR}^2}\right),
\end{align}
and we take $\beta = -1$ following~\cite{Ioffe:1981kw,Ioffe:1982ce,Haisch:2019bml}.
The parameters should be evaluated at $m_Q$ for the former contribution, 
and at the scale close to $\Lambda_\mathrm{QCD}$ for the latter contribution.
We use $m_c = 1.27\,\mathrm{GeV}$, $m_b = 4.18\,\mathrm{GeV}$, 
$\alpha_s(m_c) = 0.38$ and $\alpha_s(m_b) = 0.223$~\cite{ParticleDataGroup:2020ssz}.
The light quark masses (in the $\overline{\mathrm{MS}}$ scheme) 
also depend on the energy scale, and we take $m_u(m_c) = 2.5\,\mathrm{MeV}$
$m_d(m_c) = 5.4\,\mathrm{MeV}$, $m_u(m_b) = 1.8\,\mathrm{MeV}$ and $m_d(m_b) = 4.0\,\mathrm{MeV}$ that 
we obtain by running the light quark masses at $2\,\mathrm{GeV}$ following~\cite{ParticleDataGroup:2020ssz}.
Finally we take $\alpha_s = 0.5$, $\Lambda_\mathrm{IR} = 300\,\mathrm{MeV}$ 
and $M = 800\,\mathrm{MeV}$ for the QCD sum rule estimation for definiteness.
For these values, the contribution from the $CP$-odd photon-gluon operator is larger by
a factor of $16$ and $10$ for the charm and bottom quarks, respectively.
In the bottom quark case, the explicit quark mass suppression is compensated by the running of the strong coupling,
and the $CP$-odd photon-gluon operator is still larger than the light quark EDM 
even with the relative suppression factor $1/m_b^2$.
By requiring that $\vert d_n \vert < 1.8\times 10^{-26}\,e\,\mathrm{cm}$, we obtain
\begin{align}
	\abs{d_c} &< 6\times 10^{-22}\,e\,\mathrm{cm},
\end{align}
for the charm quark, and
\begin{align}
	\abs{d_b} &< 2\times 10^{-20}\,e\,\mathrm{cm},
\end{align}
for the bottom quark, respectively.

The constraint from $d_n$ is stronger than that from $C_S$ by more than an order of magnitude.
However, one should note that the estimation of $d_n$ has a large hadronic uncertainty.
Indeed, the final result is affected by a factor two
if we use the sum rule of the nucleon kinetic term instead of the mass term for $\lambda_n$.
There are also uncertainties related to the choice of $M$ and $\beta$ which can again result
in a factor of a few difference in the final result.
Therefore our constraint here should be understood as an order-of-magnitude estimation 
and the numerical factor should be taken with care.
In contrast, our calculation of $C_S$ is far more precise.
Its uncertainty is estimated to be $\sim 10\,\%$ and can be further reduced if needed (see the end of Sec.~\ref{sec:CS}).
In this sense, the constraints from $d_n$ and $C_S$ are complementary to each other;
$d_n$ puts a stronger constraint on the heavy quark EDM, 
while the calculation of $C_S$ is cleaner and its uncertainty is well under control.

Our constraint on $d_c$ is stronger than~\cite{Gisbert:2019ftm} by a factor two,
while the one on $d_b$ is weaker by a factor two.
However, as we noted in the introduction and the end of Sec.~\ref{sec:CS},
our constraint directly applies to $d_c(m_c)$ and $d_b(m_b)$,
and is independent of the one in~\cite{Gisbert:2019ftm}.

\section{Conclusion}
\label{sec:conclusion}
In this paper, we have derived indirect limits on the charm and bottom quark EDMs.
The charm and bottom quark EDMs generate the $CP$-odd photon-gluon operators
and the light quark EDMs after integrating out the charm and bottom quarks.
Photon-gluon operators contribute to the semi-leptonic $CP$-odd operator $C_S$ (and ultimately to paramagnetic AMO EDMs) as well as to the neutron EDM at a non-perturbative level. 
Quark EDM dominantly contributes to the neutron and nuclear EDMs.
Performing our evaluation and using the current limits, we obtain 
\begin{align}
	\abs{d_c} &< 1.3\times 10^{-20}\,e\,\mathrm{cm},
	\quad
	\abs{d_b} < 7.6\times 10^{-19}\,e\,\mathrm{cm},
\end{align}
from the paramagnetic EDM experiments, and
\begin{align}
	\abs{d_c} &< 6\times 10^{-22}\,e\,\mathrm{cm},
	\quad
	\abs{d_b} < 2\times 10^{-20}\,e\,\mathrm{cm},
\end{align}
from the neutron EDM experiment, respectively.
Although the constraint from the neutron EDM is stronger, it has a larger hadronic uncertainty.
The uncertainty of the constraint from the paramagnetic EDM is estimated as 10\,\% and
can be improved if needed, while the uncertainty from the neutron EDM can be a factor of a few.
Our constraint is independent of the one given in~\cite{Gisbert:2019ftm} in the sense that our constraint
directly applies to the EDM operators at the quark mass threshold.
By assuming a simple scaling of $d_Q/e \propto (\alpha/\pi) m_Q/\Lambda^2_Q$,
we may translate our constraint as a lower bound on $CP$-odd new physics scale:
$\Lambda_c > 70\,\mathrm{GeV}$ and $\Lambda_b > 20\,\mathrm{GeV}$ 
from the paramagnetic EDM experiment,
and $\Lambda_c > 300\,\mathrm{GeV}$ 
and $\Lambda_b > 100\,\mathrm{GeV}$ from the neutron EDM experiment.

Our result provides an important benchmark to overcome for the LHC based measurements of
the charmed baryon EDMs~\cite{Baryshevsky:2016cul,Botella:2016ksl,Fomin:2017ltw,Bagli:2017foe,Aiola:2020yam}. The idea of using the bent crystal technique for studying electromagnetic properties of baryons containing a heavy quark is very appealing. However, given the strength of the bounds derived in our work, and the necessity to satisfy independent constraints from $d_n$ and $C_S$ (hence removing a chance of accidentally large $d_{c(b)}$ due to cancellations), one may want to re-evaluate the main goal of the charmed baryon experiment. While it will be difficult to match the indirect sensitivity to $d_{c(b)}$, 
the planned measurement may achieve sufficient accuracy to extract the values of the magnetic moments $\mu_{c(b)}$ and compare it with the QCD predictions.

\paragraph{Acknowledgements}
Y.E. and M.P. are supported in part by U.S.\ Department of Energy Grant No.~de-sc0011842. M.P. would like to thank Dr. K. Melnikov for the advice in evaluating loop contributions. 
The Feynman diagrams in this paper are generated by \texttt{TikZ-Feynman}~\cite{Ellis:2016jkw}.

\appendix

\section{Convention}
\label{app:convention}

Here we summarize our conventions used in this paper.
The field strengths are given by
\begin{align}
	F_{\mu\nu} &= \partial_\mu A_\nu - \partial_\nu A_\mu,
	\quad
	G_{\mu\nu}^a = \partial_\mu G_\nu^a - \partial_\nu G^a_\mu + g_s f^{abc}G_\mu^b G_\nu^c,
	\quad
	G_{\mu\nu} = G_{\mu\nu}^a T^a,
\end{align}
where $f^{abc}$ is the SU(3) structure constant.
The SU(3) generator satisfies 
\begin{align}
	\left[T^a, T^b\right] = if^{abc}T^c,
	\quad
	\mathrm{tr}_\mathrm{c}\left[T^a T^b\right] = \frac{\delta^{ab}}{2}.
\end{align}
The dual  field strengths are defined as
\begin{align}
	\tilde{F}^{\mu\nu} = \frac{1}{2}\epsilon^{\mu\nu\rho\sigma}F_{\rho\sigma},
	\quad
	\tilde{G}^{\mu\nu} = \frac{1}{2}\epsilon^{\mu\nu\rho\sigma}G_{\rho\sigma},
\end{align}
with $\epsilon^{0123} = +1$.
We take the gamma matrix as
\begin{align}
	\left\{ \gamma^\mu, \gamma^\nu\right\} = 2\eta^{\mu\nu},
	\quad
	\eta_{\mu\nu} = \mathrm{diag}(+, -, -, -),
	\quad
	\gamma_5 = i\gamma^0 \gamma^1 \gamma^2 \gamma^3 = 
	- \frac{i}{4!}\epsilon^{\mu\nu\rho\sigma}\gamma_\mu \gamma_\nu \gamma_\rho\gamma_\sigma,
\end{align}
and $\sigma_{\mu\nu}$ as
\begin{align}
	\sigma_{\mu\nu} = \frac{i}{2}\left(\gamma_\mu \gamma_\nu - \gamma_\nu \gamma_\mu\right)
	= \frac{i}{2}\left[\gamma_\mu, \gamma_\nu\right].
\end{align}
The charge conjugation matrix $\mathcal{C}$ satisfies
\begin{align}
	\left(\gamma^\mu\right)^{T} \mathcal{C} = - \mathcal{C}\gamma^\mu,
\end{align}
such that $\psi^{T}\mathcal{C}$ has the same Lorentz transformation property as $\bar{\psi}$.
In the Weyl representation it is given by
\begin{align}
	\mathcal{C} = i\gamma^0 \gamma^2.
\end{align}

\section{Technical details}
\label{app:details}
In this appendix we provide some technical details that we omit in the main text.

\subsection{Derivation of $CP$-odd photon-gluon operators}
\label{app:photon-gluon}
In this subsection we provide derivation of the $CP$-odd photon-gluon operators~\eqref{eq:Leff_photon-gluon}
after integrating out the heavy quarks (see~\cite{Novikov:1983gd} for an extensive review on this procedure).
Our starting point is Eq.~\eqref{eq:effective_action_dQ}.
We can simplify it as
\begin{align}
	S_\mathrm{eff} &= - \frac{id_Q}{2}\mathrm{Tr}
	\left[\frac{i\slashed{D} + m_Q}{i\slashed{D} + m_Q}\frac{1}{i\slashed{D} - m_Q}
	\sigma_{\mu\nu}\tilde{F}^{\mu\nu} \right]
	= -\frac{i d_Q m_Q}{2}\mathrm{Tr}
	\left[\frac{1}{ \left(i D\right)^2 - m_Q^2 + \frac{1}{2}\sigma \cdot H} (\sigma\cdot \tilde{F})
	\right],
\end{align}
where $(iD)^2 = iD_\mu iD^\mu$ and we used that the traces of odd $\gamma$'s vanish in the second equality.
We further expand the denominator with respect to $H$,
and obtain up to fourth-order in the gauge fields
\begin{align}
	S_\mathrm{eff} &=
	- \frac{id_Q m_Q}{2}\left(T_2 + T_3 + T_4\right),
\end{align}
where
\begin{align}
	T_2 &= -\frac{1}{2}\mathrm{Tr}\left[
	\frac{1}{ \left(i D\right)^2 - m_Q^2} (\sigma \cdot H) \frac{1}{ \left(i D\right)^2 - m_Q^2} 
	(\sigma\cdot \tilde{F})
	\right], \\ 
	T_3 &= +\frac{1}{4}\mathrm{Tr}\left[
	\frac{1}{\left(i D\right)^2 - m_q^2} 
	(\sigma \cdot H)  \frac{1}{\left(i D\right)^2 - m_Q^2} 
	(\sigma \cdot H)  \frac{1}{\left(i D\right)^2 - m_Q^2} 
	(\sigma\cdot \tilde{F})
	\right], \\
	T_4 &= -\frac{1}{8}\mathrm{Tr}\left[
	\frac{1}{ \left(i D\right)^2 - m_Q^2}(\sigma \cdot H)  \frac{1}{\left(i D\right)^2 - m_Q^2} 
	(\sigma \cdot H)  \frac{1}{\left(i D\right)^2 - m_Q^2} 
	(\sigma \cdot H)  \frac{1}{\left(i D\right)^2 - m_Q^2} 
	(\sigma\cdot \tilde{F})
	\right].
\end{align}
The term linear in $\sigma_{\mu\nu}$ vanishes after taking the trace of the spinor index.
In order to compute $T_2$, we use the following identity~\cite{Novikov:1983gd}:
\begin{align}
	\mathrm{Tr}\left[\frac{1}{\left((iD)^2 - m_q^2\right)^3}f(H, F)\right]
	&= \mathrm{Tr}\left[\frac{1}{\left((i\partial)^2 - m_q^2\right)^3}f(H, F)\right]
	- \mathrm{Tr}\left[\frac{1}{\left((iD)^2 - m_q^2\right)^5} H_{\mu\nu}H^{\mu\nu} f(H, F)\right],
	\label{eq:identity_one_loop}
\end{align}
where $f(H, F)$ is an arbitrary function of $H_{\mu\nu}$ and $F_{\mu\nu}$.
We thus obtain
\begin{align}
	\frac{d}{dm_Q^2}T_2 &=
	-\mathrm{Tr}\left[\frac{1}{((i\partial)^2 - m_Q^2)^3} \left(\sigma \cdot H\right)\sigma \cdot \tilde{F} \right]
	+ \mathrm{Tr}\left[\frac{1}{((iD)^2 - m_Q^2)^5}
	H_{\mu\nu}H^{\mu\nu}\left(\sigma \cdot H\right) \sigma \cdot \tilde{F} \right].
\end{align}
where we ignore the derivatives acting on $F$.
We can now replace $iD$ by $i\partial$ in the last term to the order of our interest, and obtain
\begin{align}
	\frac{d}{dm_Q^2}T_2 &= -i\int d^4 x \left[ 
	\frac{3eQ_Q}{4\pi^2 m_Q^2} F_{\mu\nu}\tilde{F}^{\mu\nu}
	+ \frac{1}{24\pi^2 m_Q^6} \mathrm{tr}_c \left[H_{\mu\nu}H^{\mu\nu} H_{\rho\sigma}\tilde{F}^{\rho\sigma}\right]
	\right].
\end{align}
By integrating this we obtain
\begin{align}
	T_2 &= i\int d^4 x \left[ 
	\frac{3eQ_Q}{4\pi^2} \ln\left(\frac{M_R^2}{m_Q^2}\right) F_{\mu\nu}\tilde{F}^{\mu\nu}
	+ \frac{1}{48\pi^2 m_Q^4}\mathrm{tr}_c \left[H_{\mu\nu}H^{\mu\nu} H_{\rho\sigma}\tilde{F}^{\rho\sigma}\right]
	\right],
\end{align}
where the additional mass scale $M_R$ comes from the regularization which we did not write down explicitly.
Next we compute $T_3$. By ignoring the derivatives acting on $\tilde{F}_{\mu\nu}$, we obtain
\begin{align}
	T_3 = \frac{1}{4}\mathrm{Tr}&\left[
	\frac{1}{((iD)^2 - m_Q^2)^3} \left(\sigma \cdot H\right)^2 (\sigma \cdot \tilde{F})
	+ \frac{1}{((iD)^2 - m_Q^2)^4}\left[(iD)^2, \sigma\cdot H\right] (\sigma \cdot H) (\sigma \cdot \tilde{F})
	\right. \nonumber \\ &\left.
	- \frac{1}{((iD)^2 - m_Q^2)^5} \left[(iD)^2, \sigma\cdot H\right]\left[(iD)^2, \sigma\cdot H\right] \sigma \cdot \tilde{F}
	+ \cdots
	\right],
	\label{eq:T3}
\end{align}
where the dots indicate terms that contain more than two commutators 
and thus correspond to higher dimensional operators that are out of our interest.
The first term is easily seen to generate only $\mathcal{O}(H^4 \tilde{F})$ operators 
with the help of the identity~\eqref{eq:identity_one_loop}.
The second term is expanded as
\begin{align}
	\mathrm{Tr}\left[\frac{1}{((iD)^2 - m_Q^2)^4}\left[(iD)^2, \sigma\cdot H\right] 
	(\sigma \cdot H) (\sigma \cdot \tilde{F})\right]
	= \mathrm{Tr}&\left[\frac{1}{((iD)^2 - m_Q^2)^4} iD^\alpha \left[iD_\alpha, \sigma \cdot H\right] 
	(\sigma \cdot H) \sigma \cdot \tilde{F}
	\right. \nonumber \\ &\left.
	+ iD^\alpha \frac{1}{((iD)^2 - m_Q^2)^4}\left[iD_\alpha, \sigma \cdot H\right] 
	(\sigma \cdot H) \sigma \cdot \tilde{F}
	\right. \nonumber \\ &\left.
	+ \frac{1}{((iD)^2 - m_Q^2)^4}
	\left[iD^\alpha, \sigma\cdot H\right] \left[iD_\alpha, \sigma \cdot H\right] \sigma \cdot \tilde{F}
	\right].
\end{align}
The first two terms induce only higher order terms~\cite{Novikov:1983gd}, 
and thus we ignore them.
The third term in Eq.~\eqref{eq:T3} already contains three field strengths and two commutators,
and hence we can ignore any further commutators between $D$ and $H$.
By taking the angular average, we obtain
\begin{align}
	\mathrm{Tr}\left[\frac{1}{((iD)^2 - m_Q^2)^5} \left[(iD)^2, \sigma\cdot H\right]
	\left[(iD)^2, \sigma\cdot H\right] \sigma \cdot \tilde{F}\right]
	&= \mathrm{Tr}\left[
	\frac{(iD)^2}{((iD)^2 - m_Q^2)^5} \left[iD^\alpha, \sigma\cdot H\right]
	\left[iD_\alpha, \sigma \cdot H\right] \sigma \cdot \tilde{F} \right].
\end{align}
By combining them, we obtain
\begin{align}
	T_3 &= -\frac{m_Q^2}{4}\mathrm{Tr}\left[
	\frac{1}{((iD)^2 - m_Q^2)^5} \left[iD^\alpha, \sigma\cdot H\right]
	\left[iD_\alpha, \sigma \cdot H\right] \sigma \cdot \tilde{F}
	\right]
	\nonumber \\
	&= - \frac{g_s^2}{24\pi^2 m_Q^4}
	\int d^4x\,\mathrm{tr}_c \left[(\mathcal{D}^\alpha G_{\mu\rho})
	(\mathcal{D}_\alpha {G_{\nu}}^\rho)\right] \tilde{F}^{\mu\nu}
	= 0.
\end{align}
Finally we compute $T_4$. 
To the order of our interest, we can simply replace $iD$ by $i\partial$ in the denominator. 
It is then easy to see that
\begin{align}
	T_4 = -\frac{i}{48\pi^2 m_Q^4}\int d^4 x\,
	\mathrm{tr}_c\left[3H_{\mu\nu}H^{\mu\nu} H_{\rho\sigma}\tilde{F}^{\rho\sigma}
	- 4 {H^\mu}_\nu {H^\nu}_\rho {H^\rho}_\sigma {\tilde{F}^\sigma}_\mu
	\right].
\end{align}
Therefore we obtain
\begin{align}
	S_\mathrm{eff} = \frac{d_Q}{48\pi^2 m_Q^3}\int d^4 x\,
	\mathrm{tr}_c \left[- H_{\mu\nu}H^{\mu\nu} H_{\rho\sigma}\tilde{F}^{\rho\sigma}
	+ 2 {H^\mu}_\nu {H^\nu}_\rho {H^\rho}_\sigma {\tilde{F}^\sigma}_\mu
	\right],
\end{align}
where we ignored the quadratic term that is irrelevant for our purpose.
If we drop the gluons, it correctly reduces to the result in~\cite{Ema:2021jds}.

\subsection{Borel transformation and IR divergence}
\label{app:Borel_IRdiv}
In this subsection we discuss the Borel transformation of Eq.~\eqref{eq:Pi_OPE_bfrBorel}
that contains both the UV and IR divergences.
As the Borel transformation is related to the imaginary part,
it is equivalent to taking the imaginary part of
\begin{align}
	I(p^2; \epsilon_\mathrm{IR}, \epsilon_\mathrm{UV})
	&\equiv \Gamma(\epsilon_\mathrm{IR})
	\left(-\frac{\Lambda_\mathrm{IR}^2}{p^2}\right)^{-\epsilon_\mathrm{IR}}
	\Gamma(-\epsilon_\mathrm{UV}-\epsilon_{\mathrm{IR}}) 
	\left(-\frac{\Lambda_\mathrm{UV}^2}{p^2}\right)^{-\epsilon_\mathrm{UV}}.
\end{align}
The limit of this function at $ \epsilon_\mathrm{UV} \to 0,\:\epsilon_\mathrm{IR}\to 0 $ is not well defined.
For example, if the limit is taken along the line $ \epsilon_\mathrm{UV}=a\,\epsilon_\mathrm{IR} $
with fixed $a$, one would get an $ a $-dependent result:
\begin{equation}
\label{UV=a IR}
    \underset{\epsilon_{\mathrm{IR}}\rightarrow 0}{\text{lim}}I(p^2;\epsilon_{\mathrm{IR}},a\,\epsilon_{\mathrm{IR}})=-\frac{1}{2\left(1+a\right)}\left[\text{log}^2\left(-\frac{\Lambda_\mathrm{IR}^2}{p^2}\right)+2a\text{log}\left(-\frac{\Lambda_\mathrm{IR}^2}{p^2}\right)\text{log}\left(-\frac{\Lambda_\mathrm{UV}^2}{p^2}\right)+a^2\text{log}^2\left(-\frac{\Lambda_\mathrm{UV}^2}{p^2}\right)\right],
\end{equation}
where we only kept the double logarithmic terms. The purpose of this subsection is to understand the correct prescription of evaluating the imaginary part of this function.

In order to understand the correct prescription, it is helpful to consider a simpler example:
a scalar three-body decay.
Indeed, the two point correlator $\Pi(p)$ contains three quark propagators,
and thus the imaginary part of this function is related a three-body decay phase space integral.
Therefore we consider the following Lagrangian
\begin{align}
	\mathcal{L} = \frac{1}{2}\left(\partial \phi\right)^2 - \frac{m_\phi^2}{2}\phi^2
	+ \frac{1}{2}\left(\partial \chi \right)^2 - \frac{m_\chi^2}{2}\chi^2 - \frac{\lambda}{6}\phi \chi^3,
\end{align}
and study the three-body decay $\phi \to 3\chi$.
The two-loop diagram is evaluated as

\begin{align}
	i \mathcal{M}_2 &= 
	\begin{tikzpicture}[baseline=(p1)]
	\begin{feynman}[inline]
		\vertex (p1);
		\vertex [right = 0.75 of p1] (v1);
		\vertex [right = 0.5 of v1] (c);
		\vertex [above = 0.5 of c] (v2);
		\vertex [below = 0.5 of c] (v4);
		\vertex [right = 0.5 of c] (v3);
		\vertex [right = 0.75 of v3] (p2);
		\diagram*{
		(p1) -- [scalar, momentum=\(\scriptstyle p\)] (v1) -- (v3) -- [scalar, momentum=\(\scriptstyle p\)] (p2),
		(v1) -- [quarter left] (v2) -- [quarter left] (v3) -- [quarter left] (v4) -- [quarter left] (v1),
		};
	\end{feynman}
	\end{tikzpicture}
	= \frac{i\lambda^2}{6}\int \frac{d^4l_1}{(2\pi)^4}\int \frac{d^4l_2}{(2\pi)^4}D_F(l_1)D_F(l_2)D_F(l_1 + l_2 - p),
\end{align}
where $D_F(p)$ is the Feynman propagator,
\begin{align}
	iD_F(p) = \frac{i}{p^2 - m_\chi^2 + i0},
\end{align}
and the solid line is $\chi$ while the dashed line is $\phi$.
The cutting rule tells us that
\begin{align}
	\mathrm{Im} \mathcal{M}_2
	&=
	\frac{1}{2}\frac{\lambda^2}{6}\int d\Pi_\mathrm{LIPS},
	\label{eq:cutting_rule}
\end{align}
where $d\Pi_\mathrm{LIPS}$ is the three-body Lorentz invariant phase space integral.
This is expanded with respect to $m_\chi^2/m_\phi^2$ as
\begin{align}
	\int d\Pi_\mathrm{LIPS} &=
	 \frac{1}{32\pi^3}\left[\frac{m_\phi^2}{8} 
	+ \frac{3m_\chi^2}{4} \left(\log\left(\frac{4m_\chi^2}{m_\phi^2}\right) - 1\right) + \cdots \right],
	\label{eq:m_expansion}
\end{align}
where we used $p^2 = m_\phi^2$.
We now evaluate the left hand side of Eq.~\eqref{eq:cutting_rule}
in an analogous way as the main text.
We may expand the propagator as
\begin{align}
	iD_F(p) = \frac{i}{p^2 - m_\chi^2} = i\left(\frac{1}{p^2} + \frac{1}{p^2} m_\chi^2 \frac{1}{p^2} + \cdots \right),
\end{align}
where we omit $i\epsilon$ for notational simplicity.
In the coordinate space this is given by
\begin{align}
	iD_F(x) &= -\frac{1}{4\pi^2 x^2} - \frac{m_\chi^2}{16\pi^2}\Gamma(\epsilon_\mathrm{IR}) 
	\left(-\Lambda_\mathrm{IR}^2 x^2\right)^{-\epsilon_\mathrm{IR}} + \cdots.
\end{align}
The first order term in $m_\chi^2$ is IR divergent 
and this is analogous to our propagator with the $CP$-odd operator insertion in Sec.~\ref{subsec:QCDcondensate}.
The two-loop amplitude is given by
\begin{align}
	i\mathcal{M}_2(p) &= \frac{\left(i\lambda\right)^2}{6}\mu_\mathrm{UV}^{d_\mathrm{UV}-4} \int d^{d_\mathrm{UV}}x\,
	e^{ip\cdot x} iD_F(x) iD_F(x) iD_F(x).
\end{align}
To the first order in $m_\chi^2$ we obtain
\begin{align}
	\left.\mathcal{M}_2\right\vert_{\mathcal{O}(m_\chi^2)} &= 
	-\frac{\lambda^2 m_\chi^2}{512 \pi^4}I(p^2; \epsilon_\mathrm{IR}, \epsilon_{\mathrm{UV}}).
\end{align}
Note that we get exactly the same function $I(p^2; \epsilon_\mathrm{IR}, \epsilon_{\mathrm{UV}})$ here.
Eq.~\eqref{eq:cutting_rule} tells us that
\begin{align}
	\left.\mathrm{Im} \mathcal{M}_2\right\vert_{\mathcal{O}(m_\chi^2)} &=
	\left[\frac{1}{2}\frac{\lambda^2}{6}\int d\Pi_\mathrm{LIPS}\right]_{\mathcal{O}(m_\chi^2)}
	= \frac{\lambda^2 m_\chi^2}{512 \pi^3} \log\left(\frac{m_\chi^2}{m_\phi^2}\right),
\end{align}
where we focus on the leading logarithmic term on the right hand side.
Thus the correct prescription of evaluating the imaginary part of $I$ is
\begin{align}
	\mathrm{Im}\left[I(p^2; \epsilon_\mathrm{IR}, \epsilon_{\mathrm{UV}}) \right]
	&= -\pi \log\left(\frac{\Lambda_\mathrm{IR}^2}{p^2}\right),
\end{align}
where we identify $\Lambda_\mathrm{IR}^2 = m_\chi^2$ in the present case.
This agrees with the Borel transformation formula in~\cite{Haisch:2019bml}.
For the example in Eq.(\ref{UV=a IR}), this prescription is equivalent to neglecting the real part of the UV logarithm.

Our discussion clarifies the physical meaning of 
$\Lambda_\mathrm{IR}$ that appears in Eq.~\eqref{eq:dn_Borel_OPE}.
This IR divergence originates from the phase space integral and is regulated by the mass of the daughter particles.
In the neutron EDM case, the daughter particles are the constituent quarks.
Therefore $\Lambda_\mathrm{IR}$ is identified with the mass of the constituent up and down quarks 
that we take $\Lambda_\mathrm{IR} = 300\,\mathrm{MeV}$ in the main text.

\small
\bibliographystyle{utphys}
\bibliography{ref}
  
\end{document}